\documentclass[8.5pt,twoside,twocolumn]{article}
\usepackage[super,sort&compress,comma]{natbib} 
\usepackage{mhchem}
\usepackage{times,mathptmx}
\usepackage{sectsty}
\usepackage{balance} 

\usepackage{graphicx} %eps figures can be used instead
\usepackage{lastpage}
\usepackage[format=plain,justification=raggedright,singlelinecheck=false,font=small,labelfont=bf,labelsep=space]{caption} 
\usepackage{fancyhdr}
\usepackage[author={SHN}]{pdfcomment}

\usepackage{dblfloatfix}

\usepackage[utf8]{inputenc}
\usepackage[amssymb]{SIunits}
\usepackage{amssymb,amsmath}
\usepackage{notes2bib}
\usepackage{booktabs}
\usepackage{multirow}

\usepackage{xspace}
\newcommand{\SLcam}{SLcam\textsuperscript{\textregistered}\xspace}

\usepackage{xcolor,colortbl}
\definecolor{Gray}{gray}{0.85}

\begin{document}

\thispagestyle{plain}
\fancypagestyle{plain}{
%\fancyhead[L]{\includegraphics[height=8pt]{headers/LH}}
%\fancyhead[C]{\hspace{-1cm}\includegraphics[height=20pt]{headers/CH}}
%\fancyhead[R]{\includegraphics[height=10pt]{headers/RH}\vspace{-0.2cm}}
\renewcommand{\headrulewidth}{1pt}
}
\renewcommand{\thefootnote}{\fnsymbol{footnote}}
\renewcommand\footnoterule{\vspace*{1pt}% 
\hrule width 3.4in height 0.4pt \vspace*{5pt}} 
\setcounter{secnumdepth}{5}

\makeatletter 
\def\subsubsection{\@startsection{subsubsection}{3}{10pt}{-1.25ex plus -1ex minus -.1ex}{0ex plus 0ex}{\normalsize\bf}} 
\def\paragraph{\@startsection{paragraph}{4}{10pt}{-1.25ex plus -1ex minus -.1ex}{0ex plus 0ex}{\normalsize\textit}} 
\renewcommand\@biblabel[1]{#1}            
\renewcommand\@makefntext[1]% 
{\noindent\makebox[0pt][r]{\@thefnmark\,}#1}
\makeatother 
\renewcommand{\figurename}{\small{Fig.}~}
\sectionfont{\large}
\subsectionfont{\normalsize} 

\fancyfoot{}
%\fancyfoot[LO,RE]{\vspace{-7pt}\includegraphics[height=9pt]{headers/LF}}
%\fancyfoot[CO]{\vspace{-7.2pt}\hspace{12.2cm}\includegraphics{headers/RF}}
%\fancyfoot[CE]{\vspace{-7.5pt}\hspace{-13.5cm}\includegraphics{headers/RF}}
\fancyfoot[RO]{\footnotesize{\sffamily{1--\pageref{LastPage}~\textbar  \hspace{2pt}\thepage}}}
\fancyfoot[LE]{\footnotesize{\sffamily{\thepage~\textbar~%\hspace{3.45cm} 
 1--\pageref{LastPage}}}}
\fancyhead{}
\renewcommand{\headrulewidth}{1pt} 
\renewcommand{\footrulewidth}{1pt}
\setlength{\arrayrulewidth}{1pt}
\setlength{\columnsep}{6.5mm}
\setlength\bibsep{1pt}

\twocolumn[
  \begin{@twocolumnfalse}
\noindent\LARGE{\textbf{%
Road to micron resolution with %the full-field energy dispersive 
 a color X-ray camera 
-- polycapillary optics characterization
}}
\vspace{0.6cm}

\noindent\large{\textbf{%
Stanisław H. Nowak,$^{\ast}$\textit{$^{ab}$} 
Marko Petric,\textit{$^{cd}$} 
Josef Buchriegler,\textit{$^{d}$} 
Aniouar  Bjeoumikhov,\textit{$^{af}$}
Zemfira  Bjeoumikhov,\textit{$^{a}$}
Johannes von Borany,\textit{$^{d}$} 
Frans Munnik,\textit{$^{d}$} 
Martin Radtke,\textit{$^{g}$}  
Axel D. Renno,\textit{$^{h}$} 
Uwe Reinholz,\textit{$^{g}$} 
Oliver Scharf,\textit{$^{a}$} 
Joachim Tilgner,\textit{$^{a}$}
and 
Reiner Wedell\textit{$^{i}$} 
}}\vspace{0.5cm}
%Please note that \ast indicates the corresponding author(s) but no footnote text is required. 

%\noindent\textit{\small{\textbf{Received Xth XXXXXXXXXX 20XX, Accepted Xth XXXXXXXXX 20XX\newline
%First published on the web Xth XXXXXXXXXX 200X}}}
%
%\noindent \textbf{\small{DOI: 10.1039/b000000x}}
%\vspace{0.6cm}
%%Please do not change this text.

\noindent \normalsize{%
%The color X-ray camera \SLcam is a full-field, single photon detector providing scanning free, energy and spatial resolved X-ray imaging. 
%
% on a pnCCD. 
%The use of a subpixel resolution algorithm allows % signal from individual capillary channels can be distinguished   
%a release of \SLcam spatial resolution from pixel size influence  confining it to the diameter of individual capillary channels.
In a color X-ray camera spatial resolution is achieved by means of a polycapillary optic conducting X-ray photons from small regions on a sample to distinct energy dispersive pixels on a CCD matrix.
At present, the resolution limit of color X-ray camera systems can go down to several microns and is mainly restricted by pixel dimensions. % and diameter of polycapillary channels.
The recent development of an efficient subpixel resolution algorithm allows % signal from individual capillary channels can be distinguished   
a release from pixel size, limiting the  resolution only to the quality of the optics. 
In this work 
 polycapillary properties  that influence the spatial resolution  are systematized and assessed both theoretically and experimentally. 
It is demonstrated that with the current technological level reaching one micron resolution is challenging, but possible.
}
\vspace{0.5cm}
 \end{@twocolumnfalse}
  ]

%Footnotes
%\footnotetext{\dag~Electronic Supplementary Information (ESI) available: [details of any supplementary information available should be included here]. See DOI: 10.1039/b000000x/}

%Please use \dag to cite the ESI in the main text of the article.
%If you article does not have ESI please remove the the \dag symbol from the title and the above footnotetext.

\footnotetext{\textit{$^{a}$~IfG -- Institute for Scientific Instruments GmbH, Berlin, Germany.
%Fax: +49 30 6392-6501;
%Tel: +49 30 6392-6504;
%E-mail: nowak@ifg-adlershof.de
%Address, Address, Town, Country. Fax: XX XXXX XXXX; Tel: XX XXXX XXXX; E-mail: xxxx@aaa.bbb.ccc
}}
\footnotetext{\textit{$^{b}$~Stanford Synchrotron Radiation Lightsource,
		Menlo Park, California, USA
		%Fax: +49 30 6392-6501;
		%Tel: +49 30 6392-6504;
		%E-mail: nowak@ifg-adlershof.de
		%Address, Address, Town, Country. Fax: XX XXXX XXXX; Tel: XX XXXX XXXX; E-mail: xxxx@aaa.bbb.ccc
	}}
\footnotetext{\textit{$^{c}$~J. Stefan Institute, Ljubljana, Slovenia.}}
\footnotetext{\textit{$^{c}$~University of Zagreb, Faculty of Geotechnical Engineering, Hallerova aleja 7, Varazdin, HR - 42 000.}}
\footnotetext{\textit{$^{e}$~Helmholtz-Zentrum Dresden-Rossendorf, Dresden, Germany. }}
\footnotetext{\textit{$^{f}$~Institute for Computer Science and Problems of Regional Management,
 Kabardino-Balkaria, Russia. }}
\footnotetext{\textit{$^{g}$~BAM Federal Institute for Material Research and Testing, Berlin, Germany.}}
\footnotetext{\textit{$^{h}$~Helmholtz-Zentrum Dresden-Rossendorf, Helmholtz Institute Freiberg for Resource Technology,
Freiberg, Germany.}}
\footnotetext{\textit{$^{i}$~IAP Institute for Applied Photonics e.V., Berlin, Germany.}}
%additional addresses can be cited as above using the lower-case letters, c, d, e... If all authors are from the same address, no letter is required

\section{Introduction}

In the last decades X-ray fluorescence (XRF) underwent an  evolution from broad
area element analysis towards spatially resolved elemental imaging \cite{Beckhoff2007}.
A conventional approach to spatially resolved XRF employs a focused X-ray
beam to map the fluorescence of a given sample region \cite{Haschke2014}.
Each mapping position is reached using a fine mechanical
$xyz$-stage.
XRF spectra are achieved for each scanning position by means of energy resolved detectors.
%With a use of spatial resolving detectors a detailed map of XRF spectra is achieved.

In the last years a cross over from compact 1-D detectors to spatially resolving detectors was obtained  with the use of polycapillary optics and charge coupled devices (CCD)~\cite{Sakurai1999_Total-reflection,Sakurai2004_Fast,Alfeld2010,Romano_2016}. 
However, most of these detection systems can only measure spatial
information at a pixel size and without energy
resolution.\cite{Mail2009microscintigraphy,Sakurai2004_Fast,Eba2006_Rapid,Tsuji2015}

The recent development of \SLcam \cite{Kuehn2011,Ordavo2011250,Scharf2011} filled that gap bringing a high quantum efficiency and throughput color X-ray camera system  that combines a
spatially and spectrally resolving pn-junction Charged-Coupled Device (pnCCD) from PNSensor in Munich\cite{Strueder2001}  with polycapillary
optics~\cite{Bjeoumikhov2003} from IfG – Institute for Scientific Instruments GmbH in Berlin.

The pnCCD detector is designed for ultrafast readout allowing detection of single photons with both spatial and energy resolution. 
Single photon counting mode is achievable even at relatively high photon count rates as available at synchrotron or PIXE endstations.\cite{Nowak2015_Examples_of_XRF_and_PIXE_imaging} %[Hanf]

A standalone pnCCD is used for spatially resolved X-ray transmission\cite{Boone2014644}, diffraction\cite{Donges2013} or scattering measurements\cite{Abboud2011}. 
However, in order to achieve an image of X-ray fluorescence, 
%divergent X-ray sources,%
X-ray photons should be explicitly guided from  small regions on a sample
to corresponding pixels on a detector.
For this a polycapillary optic is used that can be regarded as a bunch of independent X-ray channels bringing the photons from the source to a proper section of a CCD,
similarly to the way a fiber optic guides light. 
The color X-ray camera uses conical and   parallel polycapillary structures. Conical polycapillary optics can be used for magnification of the image. 
Straight structures are used to obtain a 1:1 image on the detector.

The lateral resolution of  a color X-ray camera is limited by both the pixel dimension and entry diameter of a single capillary. 
Currently used \SLcam optics are optimized for  pnCCD pixel dimension of \unit{48\times48 }{\mu m^2}.\cite{Scharf2011,Ordavo2011250} 
The capillary exit diameter is adapted in such a way, that a spot size
from an individual channel on the detector is approximately equal to the pixel size.
Keeping the pixel size fixed, the spatial resolution  can be seriously improved by the use of
conically shaped magnifying optics.
%  Poly Capillary Conic Collimators (Poly-CCC).
%Single channel of Poly-CCC has a  conical shape resulting in a small entrance  to exit diameter ratio. 
%Consequently, 
In this case the signal reaching a single pixel on the detector originates from a much smaller area on a sample.  
The drawback is a limited filed of view.

With the use of subpixel algorithm the dominant role of the pixel size is  released.\cite{Nowak2015_Sub-pixel_resolution}
This algorithm  divides the signal assigned to each physical pixel over a number or virtual subpixels. 
Such an approach gives  room for further downscaling of channels, and consequently, to improvement of the spatial resolution. 
In this paper 
the polycapillary physical parameters, including but not limited to channels' diameter, are asses and
the influence of these parameters on the imaging capabilities is investigated.

\section{Spatial resolution }

% \subsection{Resolution}

Spatial resolution is the ability of an imaging device to  capture finely spaced details.
%Typically it is assessed as the maximum number of bright-dark line pairs  per unit length that can be visibly resolved with a specified contrast level. 
Typically it is assessed as $r$ -- the  maximum frequency of bright-dark line pairs in a unit length that can be visibly resolved with a specified contrast level. 
Alternatively, resolution can be describe in a reciprocal manner as resolving power or resolution limit, i.e., the shortest distance $R$ of two objects at which a  contrast above a specified limit can be maintained. 

It should be noticed than the concept of resolution is associated with a certain ambiguity.\cite{Stelzer1998}
First, the cut-of contrast level can be chosen in many, equally good manners. 
Second, in real measurements the contrast increases with the signal to noise ratio. The latter in turn is an increasing function of the image acquisition time and illumination, and can vary from one measurement to another. 
%\pdfcomment{I plan to add here a Figure showing the influence of noise and sampling.}
 	 	
\subsection{Theoretical limits to resolution}

%  Nyquist–Shannon sampling theorem

There are  some theoretical limits to spatial resolution.
The Nyquist–Shannon sampling theorem \cite{Shannon1949} says that the highest possible frequency that can be correctly reproduced from a discrete sampled signal is half of the sampling rate.
In other words, making $n$ samples over a certain distance only a signal comprising less than $n/2$ elements can be correctly distinguished. 
A higher number of elements will lead to signal aliasing resulting in distortions and artifacts. 
In the case of color X-ray camera that means that correct rendering of elements smaller than %double the pixel or 
double the capillary channel size should not be possible.

Another concept that can be used to theoretically assess the spatial resolution is  the response of an imaging system to a point source, the so-called Point Spread Function (PSF).
If the shape of the  PSF is known, then the resolution power can be calculated as $R_{PSF}$ -- the smallest distance at which the signal from two point sources of equal intensity creates a valley (lowest value of signal) lower than the contrast limit.
In particular for a system with a Gaussian type PSF, $R_{PSF}$ is always larger than the Full Width at Half Maximum (FWHM) -- a sum of two Gaussian distributions that meet at their FWHM does not crate a valley.

Concluding, the spatial resolution is limited by the maximum of two - double the sampling distance $d$, and the size of the Point Spread Function:
\begin{equation}
\label{eq:resolution_limit}
R>\max(2d,R_{PSF}).
\end{equation}
In the optimal case the sampling distance should be  smaller than a half of $R_{PSF}$. % should be somehow larger that twice the sampling distance. 
Such a spatial oversampling prevents creation of aliases and also increases the available contrast.

\subsection{Contrast Transfer Function \label{sec:CTF}}

Experimentally, the spatial resolution may be determined from the Contrast Transfer Function (CTF) that can be measured using standardized  resolution test charts.\bibnote{ISO 12233:2014
Photography -- Electronic still picture imaging -- Resolution and spatial frequency responses}
Typical test structures consist of repeated bar patterns or concentric wedges. 
CTF gives the contrast between the test chart structure and the background as a function of structure dimensions or periodicity (i.e., lines per mm). 
The spatial resolution is a point at  which the contrast drops below the chosen level.
Proper  measurement of CTF requires a long enough image acquisition time to ensure that the  noise level is below the contrast cut-off value. 

\subsection{Sub-pixel resolution}

Pixel resolution, understood as  total number of pixels   across the entire width and height of an image,
 inevitably limits the spatial resolution
%of an image gives an inevitable limitation to spatial resolution
in a way predetermined by the Nyquist-Shannon theorem. 
The details smaller than twice the pixel size cannot be rendered correctly.
In the case of color X-ray camera, however, the dominant role of  pixel size can be released with the use of a subpixel resolution algorithm.\cite{Nowak2015_Sub-pixel_resolution,Abboud2013,Kimmel2009} 
The algorithm  divides the signal assigned to each physical pixel over a number or virtual subpixels, increasing the pixel count and bypassing the Nyquist-Shannon theorem limitation.

Images with subpixel resolution are achievable
%Sub-pixel algorithms are possible
  due to
  %Such an approach is possible due 
the specific physics of  the interaction of X-ray photons  with a CCD's Si layer.
An X-ray photon that is absorbed in a fully depleted silicon layer  generates a so-called electron cloud with a number of electrons proportional to the photon energy.
Because the cloud has a nonzero area the generated electrons are deposited in several pixels closest to the photon hit location. 
With a correct reconstruction of the charge footprint the photon hit position can be estimated with a much higher precision than the pixel size. 
In the case of the \SLcam a pixel division up to  5$\times$5 subpixels can be safely applied.\cite{Nowak2015_Sub-pixel_resolution}
% Pixel division up to  5$\times$5 subpixels can be safely applied to \SLcam images.

\subsection{Spatial resolution of \SLcam }

The resolution limit of a color X-ray camera is restricted by the larger of two factors: $R_{PSF}$ --
the width of the PSF of a  polycapillary optic, or double the sampling distance.
With the use of subpixel resolution the sampling distance can be practically identified with the capillary channel diameter $d$. 
%Accordingly, t

The best performance is achieved, if
 the subpixel size $p$
is  smaller than the half of the resolution limit of a polycapillary.
In other words
%To avoid signal aliasing  
the subpixel dimension should be smaller than at least one of two factors: the channel diameter $d$ or half of  $R_{PSF}$.

\section{Polycapillary optics}

\begin{figure}
	\centering
	\includegraphics[width=1\linewidth]{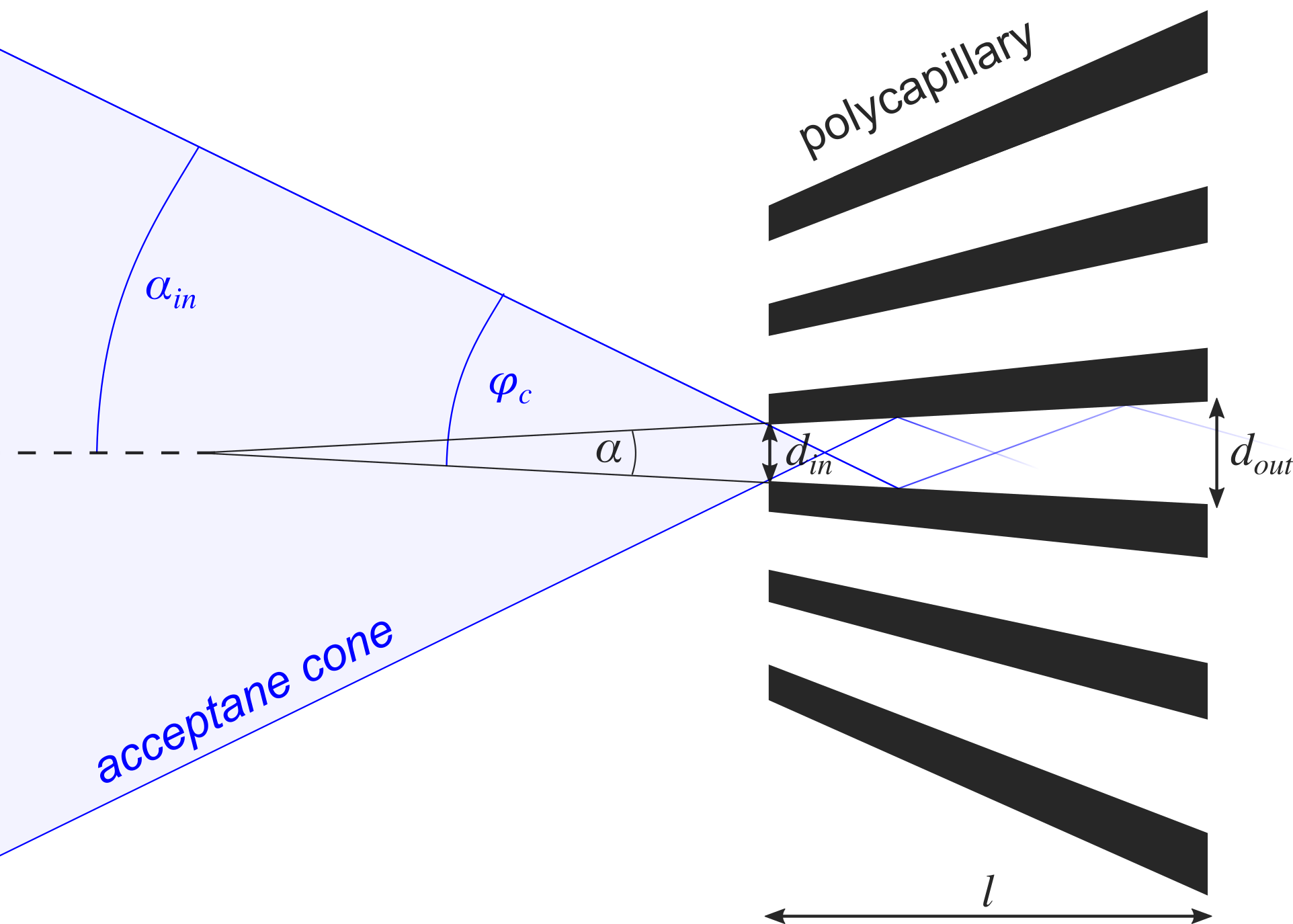}
	\caption{Schematic view of a tappered polycapillary optic.  Most relevant parameters are shown: length $l$, entrance diameter $d_{in}$, exit diameter $d_{out}$, and conical aperture $\alpha$. 
		For reference the acceptance cone with  acceptance angle  $\alpha_{in}$ is indicated with the  extreme ray paths that are reflected on the walls at the critical angle $\varphi_c$.
	}
	\label{fig:polycap}
\end{figure}

\begin{figure}
	\centering
	\includegraphics[width=.9\linewidth]{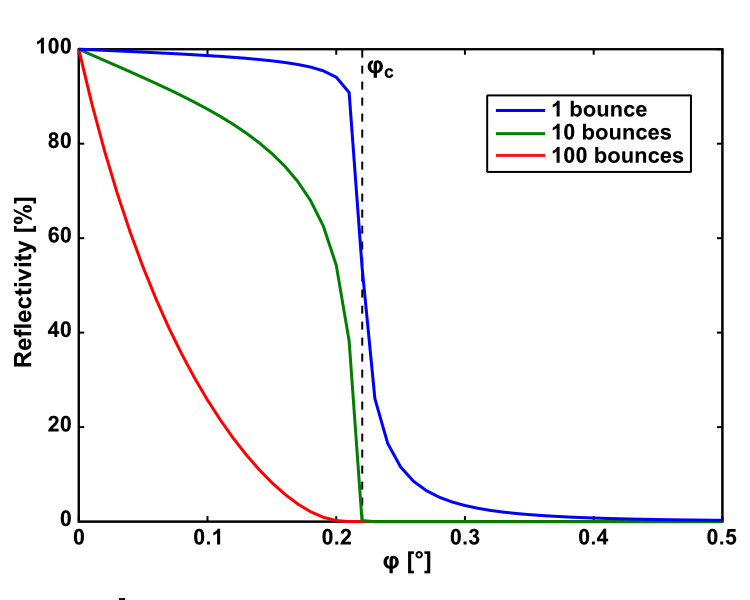}
	\caption{X-ray reflectivity in a straight capillary channel for different numbers of bounces. Calculations were performed for a glass material and X-ray energy of \unit{8}{ keV}.
		The dashed line indicates the critical angle of total reflection $\varphi_c$. 
	}
	\label{fig:reflectivity}
\end{figure}

%The pnCCD detector registers individual photons at the pixels corresponding to the hit points. Therefore a standalone detector provides imaging for measurements where spatial or angular distributions of X-ray radiation is well defined, e.g., for X-ray absorption, diffraction or scattering. For the case of X-ray fluorescence, however, radiation is emitted isotropically in the full solid angle preventing proper image formation on the detector plane. To achieve spatial resolution the detector needs to be combined with special X-ray optics that would guide photons from small regions on a sample to corresponding pixels. For that purpose a pinhole or polycapillary optics are used. 

A polycapillary optic is a bundle of glass tubes with a typical diameter of several microns separated by thin walls. For the case of imaging it can be regarded as a bunch of independent X-ray channels gathering the photons from the X-ray source and bringing them to a section of the CCD. The operating principle of polycapillary optics is based on the effect of total external reflection. Below  $\varphi_c$ -- the critical angle of total reflection, which depends mainly on the reflecting material and the X-ray photon energy, the reflection coefficient reaches values near to 100\%. In addition, the low roughness of the reflective glass surface results in a low amount  of diffuse scattered X-rays. As a result X-ray radiation is very effectively transported through an optic being reflected at the inner surface of the capillary channels. 
A schematic view of a politically optic is shown in Figure \ref{fig:polycap}.

Thanks to the  plasticity of glass, a polycapillary optic can be bent and shaped during the fabrication process.
The polycapillary objectives  are  
manufactured with either parallel and straight, 
or tapered channels. 
For the latter the entrance diameter $d_{in}$ is smaller than the exit $d_{out}$.
%In polycapillaries used for imaging, X-ray radiation
%can reach the detector 
% not only bouncing on the walls, but also directly hitting the detector.
% The importance of that is most apparent for conical optics.
 %higher energy X-rays, for which the angle of total reflection  approaches~\unit{0}{\degree}.

Straight, parallel optics are ideal for one to one imaging.  
This specific capillary geometry  results in a deep depth of sharpness, making it excellent to visualize uneven objects~\cite{Reiche2013}.
Conically shaped polycapillaries are used for optical magnification.
Magnification factors up to $M=10$ are attainable with the current fabrication technology.
%In this case  resolution is restricted by the small entry diameter that technically can reach a sub-micron scale.  
%The imaging polycapillary optics are developed with a conically shaped basic form and straight channels.

In the \SLcam the optic housing and the camera head are connected via a fine screw thread to a thin Be entrance  window of the camera, allowing  rapid changing of the X-ray lenses. 
The pnCCD chip is located \unit{6.5}{mm} below that window. 
The optic housing leaves an additional \unit{1\ to\ 2}{mm} space to prevent any damage of the fragile Be window.
All together the optic-detector clearance is about \unit{8}{mm}.

%The most important parameters characterizing the imaging polycapillary optics are: length l, field of view A, channel entrance diameter d, fill factor F, magnification factor M, acceptance angle $\alpha$, and photon collection efficiency P.

\subsection{Transmission, Acceptance,  Sensitivity}

%\begin{figure}
%\centering
%%\includegraphics[width=.9\linewidth]{cap}\\
%\includegraphics[width=.9\linewidth]{Capillary-schema_V3}
%\caption{Schematic view of a transmission trough a single capillary channel}
%\label{fig:cap}
%\end{figure}

%It can be defined as a ratio of photons emitted from a point source that are collected and transmitted by an optic to the total number of emitted photons.

%Obviously $S$ is related to optical transmission which, in turn,  
%In case of a polycapillary  structure  transparency 
The optical transmission of a polycapillary 
is strongly coupled to  its open area ratio $O$, i.e., the ratio of the polycapillary area that is not occupied by the glass to the total area.
% walls of the capillary channels and the total field of view. 
The maximum transmission $T_{max}$ of a parallel X-ray beam perfectly aligned to the polycapillary channels should not exceed $O$. 
An exception to that is the hard X-ray regime in which glass becomes transparent, allowing an X-ray beam to penetrate the whole length of polycapillary facets. 

It should be noticed that a polycapillary can only propagate photons entering within a certain acceptance cone.
The acceptance cone is a result of X-ray reflections in the channels and, in case of a  magnifying lens, also the conical shape of an optic.
The half-angle of the acceptance cone, the so called acceptance angle~$\alpha_{in}$, 
is roughly the sum of the critical angle for total reflection~$\varphi_c$ and half of the conical aperture of a channel~$\alpha$:
%is strongly dependent on the critical angle of total reflection;
%however, due to possible spiral photon propagation in a channel $\alpha$ is usually grated to the latter.
\begin{equation}
\label{eq:alpha_in}
\alpha_{in} \approx \varphi_c + \alpha/2.
\end{equation}

There are opposing causes for deviations from that sum.
First is that the probability of a photon transfer  decreases with the number of bounces of the X-ray beam. 
In Figure~\ref{fig:reflectivity} the transmission probabilities are presented as reflectivity curves for different numbers of reflections. 
For a large number of bounces the angular range for effective transmission is strongly reduced.
A not negligible surface roughness and waviness of polycapillary channels further reduce the reflectivity in consecutive bounces. 
Second effect is related to possible spiral photon propagation in a polycapillary channel.\cite{Vincze1998Optimization}
This effect enlarges the acceptance cone and is most of all related to conical optics where the number of bounces is reduced.
Another effect, related to conical polycapillaries, is the halo that occurs when X-rays penetrate  the walls between channels and are totally reflected in a neighboring channel.\cite{Proost_2003}

Obviously, a nonzero acceptance angle  leads to an increased number of photons transmitted by an optic. 
Accordingly, the parallel beam transmission is not a sufficient parameter to characterize the polycapillary's ability to collect photons. 
In this regard a much better quantity is the optical sensitivity  $S$ that  can be defined as a rate of photons emitted from a point source that are first collected and then transmitted by an optic.
In order to correctly evaluate $S$, the transmission of the radiation of a point source is integrated over the complete solid angle and divided by 4$\pi$ for normalization purpose:
% the total number of photons that are transmitted trough an optic we define the optical sensitivity $S$ as integrated transmission over the complete solid angle divided by 4$\pi$:
\begin{equation}
S= \iint {T(\phi,\vartheta) \over 4\pi} d\Omega ,
\end{equation} 
where $\phi$ and $\vartheta$ are the angular directions of the emitted radiation.
Accordingly, $S$ is the ratio between the number of photons transmitted trough the optic and the total number of photons emitted from a point source in the full solid angle.

\subsection{Point Spread Function}

The acceptance cone increases the photon collection efficiency, but also leads to a certain image blur.
%Capillary optics propagates photons that enters the polycapillary optics only within the acceptance cone. 
The point source signal arriving to a polycapillary is accepted  in a spot of a diameter $R_{in}$ which can be calculated from the acceptance angle $\alpha_{in}$ and the sample-optic distance $f_1$: 
\begin{equation}
R_{in}= 2f_1  \tan{ \alpha_{in} }.
\end{equation}

Each irradiated capillary channel transfers the photons to its exit and creates a divergent beam reaching the detector.
%This divergent beam also propagates on a cone. 
For a parallel 1:1 optic the divergent beam has the same shape as the acceptance cone.   
For the tapered, magnifying polycapillaries the half-angle of the divergence cone $\alpha_{out}$ is always smaller than $\alpha_{in}$, but also bigger than half of the conical aperture of a channel:
\begin{equation}
{\alpha\over 2} < \alpha_{out} \leq \alpha_{in}.
%\alpha_{out} \in \left(\varphi_c - {\alpha\over 2},{\alpha\over 2}\right)
\end{equation}

  The size of a single channel footprint on a detector $ R_{out} $ is a function of: $ d_{out}$ -- capillary  diameter at the exit ($ d_{out}=Md_{in}$), $\alpha_{out}$ -- polycapillary divergence angle, and  $f_2$ -- the optic-detector clearance:
\begin{equation}
R_{out}= Md_{in} + 2 f_2  \tan{ \alpha_{out} }.
\end{equation}

The PSF can be estimated as a convolution of a single channel footprint and a magnified acceptance spot. Supposing that both have a Gaussian shape, the FWHM of the resulting PSF equates to
\begin{equation}
R_{PSF}=\sqrt{{(MR_{in})^2 + R_{out} }^2 },
\end{equation}
where  $R_{in}$ is multiplied by the optic magnification factor $M$.

\subsection{Resolving power \label{sec:Resolving_power}}

According to eq. \eqref{eq:resolution_limit} $R$ -- the resolving power of  a polycapillary optic can be estimated from $R_{PSF}$ and capillary diameter $d_{in}$. 
Consequently, $R$ has its lower limit in $R_{PSF}/M$ or, following Nyquist-Shannon theorem, in $2d_{in}$:
\begin{eqnarray}
R &>& \sqrt{\left(d_{in} + 2 {f_2\over M}  \tan{ \alpha_{out} }\right)^2 + \left(2f_1  \tan{ \alpha_{in} }\right)^2}
\label{eq:R_PSF},\\
R &>& 2d_{in}.
\label{eq:R_PSF2}
\end{eqnarray}
%to be larger than $S$. 
Having such an estimate the influence of each parameter can be evaluated.
% the channel entrance diameter ($d_{in}$), magnification factor ($M$)  sample-optics ($ f_1 $) and optics-detector ($ f_2 $) distances,   angular acceptance  ($\alpha_{in}$), and divergence ($\alpha_{out}$), can be evaluated.

For the case of  nonzero  acceptance ($\alpha_{in}$) and divergence ($\alpha_{out}$), $R$ grows with the sample-optic ($ f_1 $) and optic-detector ($ f_2 $) distances. 
%By reducing these distances
%the contribution of the beam spreading can be minimized.
Typical values for $f_1$ and $f_2$ are at a level of several millimeters. 
As $\alpha_{in}$, and $\alpha_{out}$ count in milliradians, the additional spreading is in micrometer range.
It should be noted that in case of magnifying optics ($M>1$),  $f_2$ has a less significant influence on $R$.

 When  $\alpha_{out}$ and $\alpha_{in}$, or both $f_1$ and $f_2$ move toward $0$, the lower limit on resolving power, calculated with \eqref{eq:R_PSF}, approaches the capillary channel diameter $d_{in}$, {\it i.e.}, the  polycapillary sampling distance. 
 In this case the resolution is determined by  the Nyquist-Shannon theorem limitation \eqref{eq:R_PSF2}.

Finally,   $R$  decreases with  photon  energy.  
This is a consequence of the fact that both  $\alpha_{in}$ and $\alpha_{out}$  scale with the total reflection angle, which itself is  inversely  proportional to energy:
\begin{equation}
\alpha_{in} \sim \alpha_{out} \sim 1/E.
\label{eq:phi_c}
\end{equation}

%\subsection{Heterogeneity \pdfcomment{This paragraph can be deleted – it is not in the scope of this article to discuss the heterogeneity of transmission. I think this should be done in another paper that could have a title: "Road to quantitative analysis with the full-field energy dispersive color X-ray camera \SLcam"}} 
%
%An important factor when working with polycapillaries as imaging objectives is their heterogeneous structure.
%% transmission of the polycapillary structure. 
%The fabrication process of polycapillaries consist of bunching the individual channels into, usually hexagonal, bundles of several tens of elements; afterwards also the bundles are sucked together to form a structure of a larger filed of view. 
%A disadvantage of such a procedure is the variation in optical properties of individual bundles and capillary channels. 
%As a result the images acquired with polycapillary optics are affected by the heterogeneous optical sensitivity. 
%
%The X-ray images can be corrected for heterogeneous transmission, e.g., by applying Furrier transform filter.
%However, for quantitative analysis of elements and for fine image assays the images have to be corrected with carefully measured, energy dependent optical sensitivity patters. 
%
% 
\section{Experimental}

\subsection{Characterization of polycapillaries\label{sec:Characterization_of_polycapillaries}}% \pdfcomment{For the sake of consistency I should indicate the chemical composition of the glass.}

Parameters such as: magnification $M$, channel entrance diameter $d_{in 	}$,  fractional open area $O$, etc., 
%but also length of capillaries $l$ 
affect 
the performance of an optic
 altering its
  angular acceptance $\alpha_{in}$, divergence $\alpha_{out}$, and optical sensitivity $S$.
 % and, consequently,  the available  resolution limits.
In order to elucidate this influence, various polycapillaries, made of the same glass material, were investigated. 
Table \ref{tab:params} presents the comparison of physical parameters of the measured polycapillaries.
In the experiments the parallel beam transmission $T$, acceptance angle $\alpha_{in}$, and sensitivity $S$ were measured.

\begin{table}[h]
\footnotesize
\begin{tabular}{@{\extracolsep{\fill}}l c c c c c c@{\extracolsep{\fill}}}
	Optic:                          & {1:1/60}       & {1:1/24}       & {1:1/20\it ($^*\!$)} & {1:1/7}      & {8:1/2}          &  \\ \hline
	$M$ -- magnification:                 & {1}            & {1}            & {1}              & 1            & {8}              &  \\
	$d_{in}$ -- entrance\unit{}{[\mu m]}: & {60}           & {24}           & {20}             & 7            & {2}              &  \\
	$d_{out}$ -- exit\unit{}{[\mu m]}:    & {60}           & {24}           & {20}             & 7            & {16}             &  \\
	$l$ -- length\unit{}{[cm]}:           & 3              & 3              & 3 / 7 $^*$       & 3            & 8                &  \\
	$\alpha$ -- conical aperture:         & \unit{0}{\degree}    &\unit{0}{\degree}&\unit{0}{\degree}& \unit{0}{\degree} & \unit{0.1}{\degree}                 &  \\
	$A$ -- field of view\unit{}{[mm^2]}:  & {12$\times$12} & {12$\times$12} & {12$\times$12}   & {7$\times$7} & {1.5$\times$1.5} &  \\
	$O$ -- open area ratio\unit{}{[\%]}:  & 73             & 56             & 75               & 50           & 56               &
\end{tabular}
	%$\theta$ -- conical aperture:        & \unit{0}{\degree} & \unit{0}{\degree} & \unit{0}{\degree} & \unit{0.01}{\degree} &  \\
\\ \ \\
$^*$Polycapillaries 1:1/20{\it (3)} and 1:1/20{\it (7)} were produced in the same fabrication process and differ in length only. 
\caption{Comparison of  polycapillary optics discussed in the article.  
}
\label{tab:params}
\end{table}

\begin{figure}[b]
	\centering
	\includegraphics[width=1\linewidth]{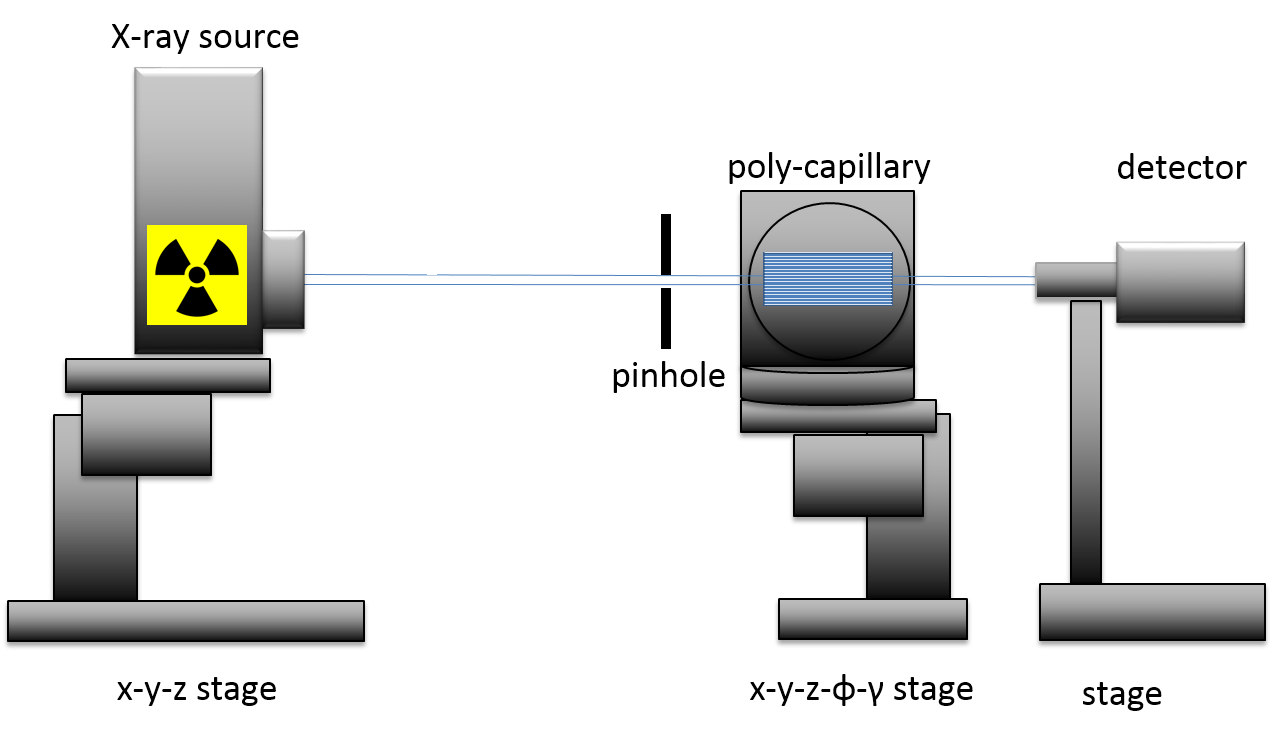}
	\caption{Setup used for transmission and angular aceptance measurements.}
	\label{fig:characterisation_setup}
\end{figure}

\begin{figure}[t]
	\centering
	\includegraphics[width=0.9\linewidth]{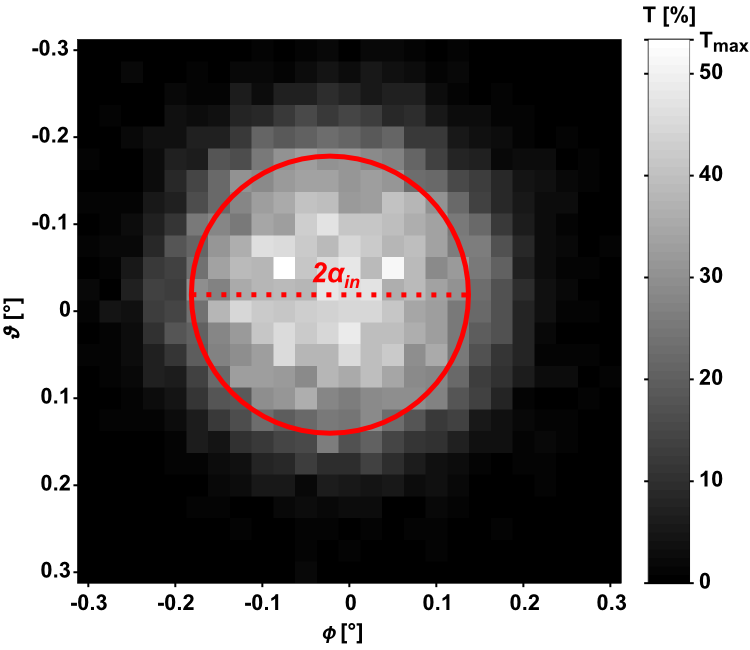}
	\caption{Exemplary angular distribution map of of the X-ray transmission obtained by scanning the angles of incoming radiation. 
		Presented is  the optic 1:1/20{\it (3)} for the X-ray range between \unit{8\ and\  10 }{keV}. 
		The red circle represents the half maximum of a fitted 2D Gaussian distribution.   
	}
	\label{fig:MAP}
\end{figure}

The experimental setup (see Figure \ref{fig:characterisation_setup}) 
comprised a micro-beam X-ray tube as an X-ray source.
A \unit{100}{\mu m} pinhole  was mounted at a large enough distance (\unit{\sim\!1}{m}) 
to achieve  an X-ray beam  with a low divergence of around %\unit{1}{m\radian} 
\unit{0.005}{\degree}. 
%(\unit{0.1}{m\radian}).
The optics were installed on a $\phi$-$\vartheta$ motorized stage allowing rotation in both directions perpendicular to the beam.
The transmitted intensity was measured by a 1D silicon drift detector (Bruker AXS Xflash Detector 430). 
The 2D angular scans were performed  with a step of  \unit{ %\Delta\phi = \Delta\vartheta =
0.025}{\degree} %(4.36~mras) 
giving  the  angular distribution maps of transmission.

Alternatively,  polycapillary parameters were  assessed with the use of \SLcam imaging.
For that purpose an image of a distant  ($d$\unit{\approx\!1}{m}) micro-beam X-ray tube was detected.
In such an arrangement the polycapillary is accessed by X-rays at various incidence angles, generating an angular map in analogy to that obtained in a scan. 
Each pixel of an \SLcam map covers an angular range of $\arctan{(p / d)}$, where $p$ is the pixel size.

%Beam intensity without a polycapillary optic was measured as a reference. 
The polycapillary optic characterization was performed in air; thus, the relevant data could only be obtained for X-ray energies above 6 keV --
the low energy part of radiation was absorbed.
In both setups beam intensity without a polycapillary optic was measured as a reference for transmission. 
The X-ray tube was to be operated at very low intensities to  avoid the creation of pileups.

The maximum of the  transmission map $T_{max}$ was identified as the  transmission of a polycapillary;
the optical sensitivity $S$ was obtained as the total sum of intensities multiplied by 
 the angular step sizes $\Delta\phi$ and $\Delta\vartheta$ and divided by $4\pi$,
\begin{eqnarray}
T_{max} &=& \max T_i\, ,\\
S&=& \sum{T_i \over 4\pi}{\Delta \phi \Delta \vartheta} \,,
\end{eqnarray}
where $T_i$ is the  value  of a single  element of the transmission map.
The angular acceptance $\alpha_{in}$ for a given energy range was calculated as the half width at half maximum of a 2D Gaussian fitted to the  map.
An exemplary  distribution map of X-ray transmission is presented in Figure \ref{fig:MAP}.

\subsection{Assessment of \SLcam resolution } 

To evaluate the resolution of the \SLcam we used  Sn Siemens star patterns  fabricated at Fraunhofer IZM in Berlin.
The Siemens star  is  deposited on a Si support and consists of concentric, evenly distributed Sn stripes 
and a little central alignment disc. 
A thin Cu layer was used to improve the  adhesion of the Sn elements.
The closer to the center of the structure, the Sn stripes are becoming thinner and the spacing between them narrower. 
At a certain point the dimensions are so little that the contrast between the structure and the background is below the predefined limit and stripes cannot be resolved. 
This is observed as a gray disk around the center of the Siemens star.

\begin{figure}
\centering
\includegraphics[width=\linewidth]{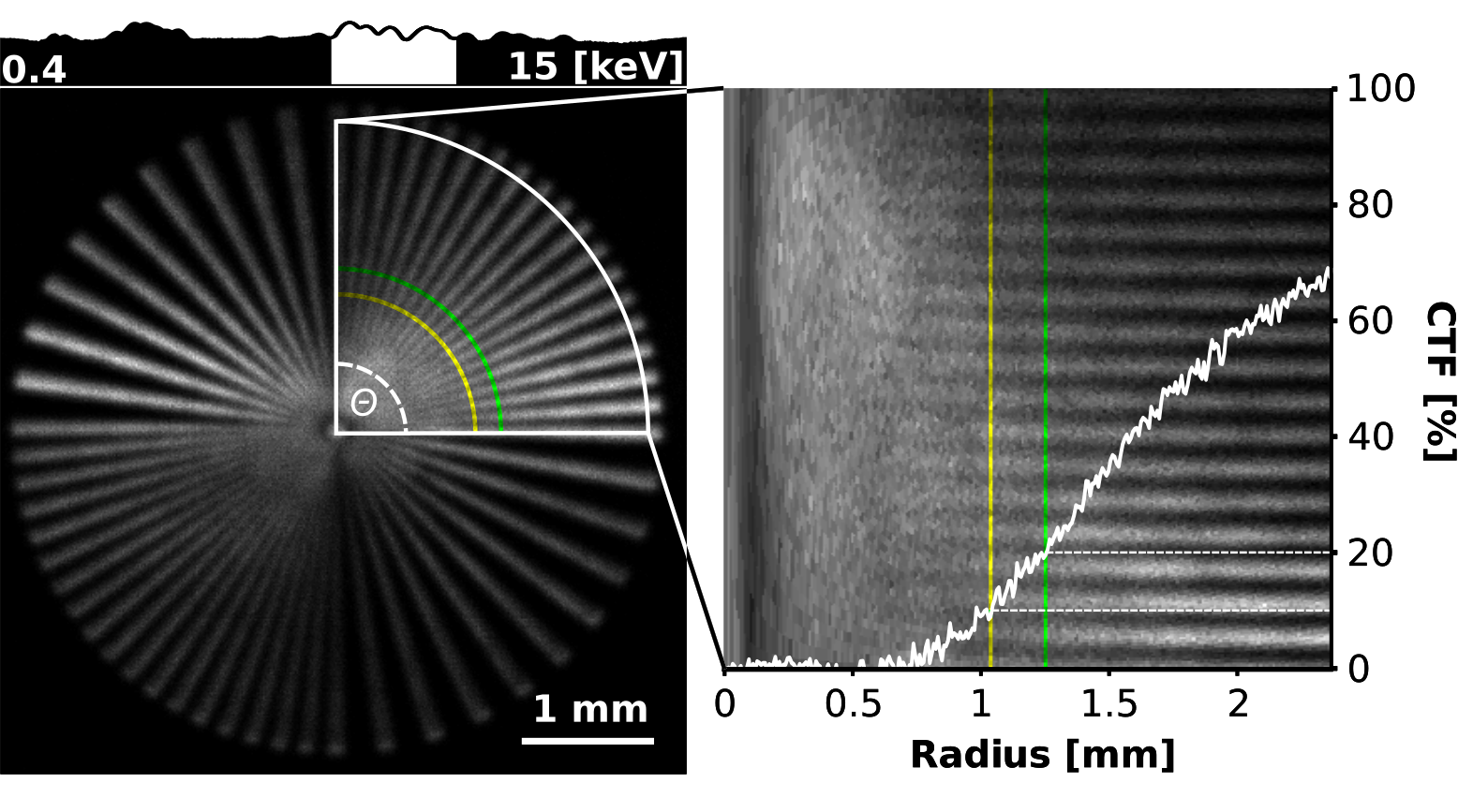}
\caption{
Assessment of the radius of the gray disk $L_\text{gray}$ for the Siemens star image.
Left: 
$4\times 4$ subpixel division image of a Siemens star with indicated circular section used for contrast transfer function (CTF) calculations.
Right: 
CTF superimposed on  the
representation of the circular section in radial coordinates.
The borders of gray rings $L_{gray}$ for contrast levels of 20\% and 10\% are indicated with green and yellow lines respectively.
The Siemens star  was measured for Cu K line series using the 1:1/7 optic; a logarithmic representation of  a sum spectrum with corresponding energy region is presented on top of the image.
}
\label{fig:CTF}
\end{figure}

The spatial resolution can be estimated from  the radius of the gray disk $L_\text{gray}$ and the number of line pairs $N_{LP}$ in a given central angle $\Theta$ of the Siemens star  (see Figure \ref{fig:CTF}).
The resolution can be given either as a maximal perceptible  line  frequency 
\begin{equation}
r={ N_{LP} \over \Theta L_\text{gray}} \;,
\end{equation}
or as resolution limit, i.e., the minimal width that is still  distinguishable:
\begin{equation}
R = { \Theta L_\text{gray} \over 2 N_{LP} } = {1 \over 2 r}\;.
\end{equation}

An accurate measure of the $L_\text{gray}$ can be assessed from CTF. 
For each value of a radius a sinusoid is fitted to an arc spanned on the the Siemens Star.
%For that purpose a sinusoid is fitted to arcs spanned on the range of $\Theta$ for the whole available range of Siemens star radii.
The value of CTF  is obtained as the ratio of the sinusoid amplitude to the average intensity on the arc.
$L_\text{gray}$ is found as the largest radius for which CTF is below the predefined contrast level.
In Figure \ref{fig:CTF} an example assessment of $L_\text{gray}$ for contrast levels of 20\% and 10\% are shown.

 \SLcam measurements were performed  employing synchrotron radiation and proton beam excitation.
 The synchrotron light was provided by the BAM{\it line} at BESSY~II.\cite{Riesemeier2005}
 The proton beam was accessed  at a newly developed High-Speed PIXE (HS-PIXE) beam line at Ion Beam Center at Helmholtz-Zentrum Dresden-Rossendorf (HZDR).~\cite{Hanf2016}
% Detailed descriptions of the two setups can be found in Ref. \citenum{Scharf2011} and \citenum{Nowak2015_Examples_of_XRF_and_PIXE_imaging}.
 
 Only valid  photon events  were selected and analyzed.\cite{Scharf2011}
 The subpixel analysis was performed with the algorithm described in ref \citenum{Nowak2015_Sub-pixel_resolution}.
 If not stated differently, presented \SLcam images were acquired at BESSY~II.

\section{Results}

 \begin{figure}
 	\centering
 	\includegraphics[width=.95\linewidth]{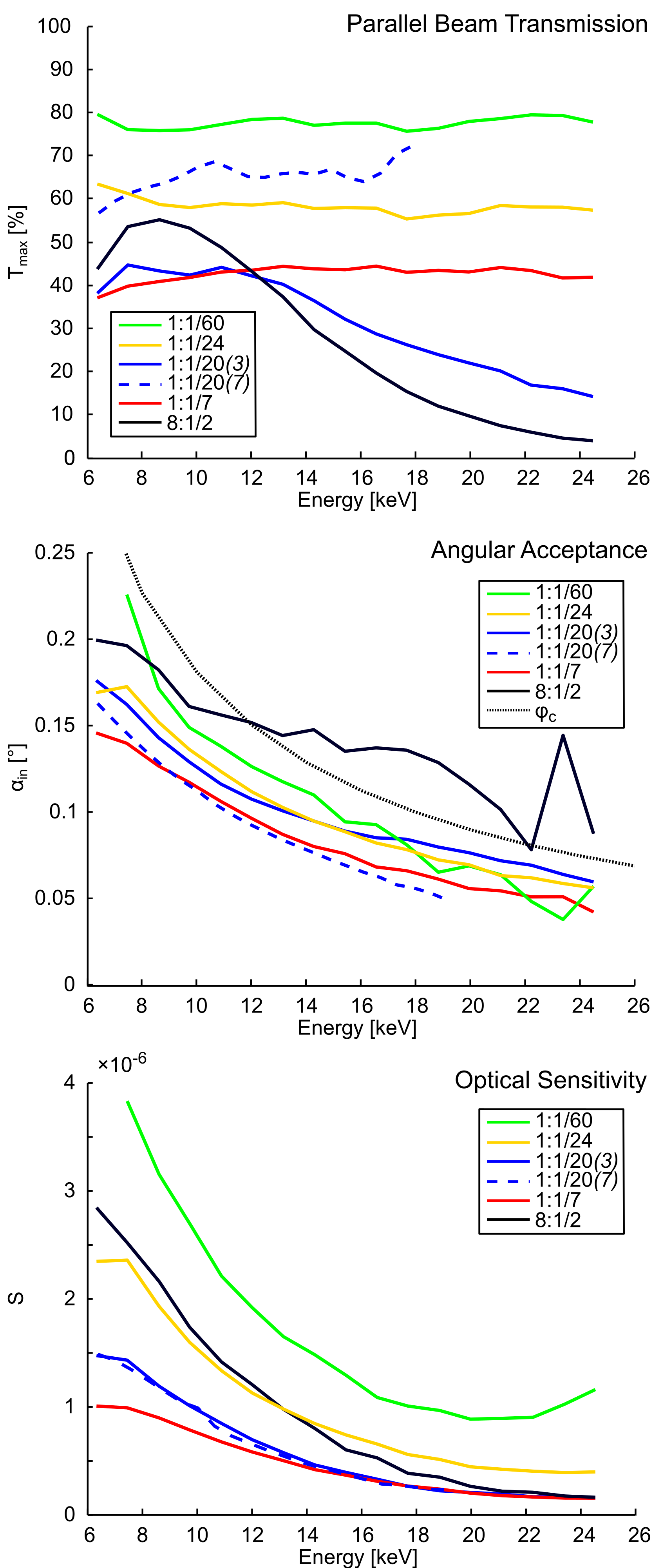}
 	\caption{Comparison of  parallel beam transmission $T$, angular acceptance angle $\alpha_{in}$, and optical sensitivity $S$ of investigated polycapillaries. 
 		%Optics 1:1/20 were found to have very similar characteristics; for the sake of legibility only curves corresponding to 1:1/20{\it(3)} are plotted. 
 	}
 	\label{fig:characterisation_curves}
 \end{figure}

\subsection{Polycapillaries}

In Figure \ref{fig:characterisation_curves}  different optics are compared for their  parallel beam transmission $T_{max}$, angular acceptance angle $\alpha_{in}$, and optical sensitivity $S$.
The parameters are plotted as a function of photon energy between \unit{6\ and\ 26}{keV}. 
For \SLcam measurements the energy range spans from \unit{6\ to\ 20}{keV}.
%Be advised that results for both 1:1/20 optics were very similar and only one set of curves is represented.

The open area ratios ($O$) listed in Table~\ref{tab:params}, correspond quite well to 
the maximal values of the parallel beam  transmission $T_{max}$.
The most significant discrepancy is seen for 1:1/20{\it(3)} optic.
As an effect of its extensive use, this polycapillary has been stained with fingerprints and dust that attenuate the X-ray radiation.
%The 1:1/20{\it(7)}  optic, has transmission  much closer to $O$.  
For some other optics (1:1/60 and 1:1/24) $T_{max}$ is slightly overestimated. 
This overestimation seems to be a result of a noise sensitive quantification procedure that selects the most intense pixel from a transmission map (see 
Section \ref{sec:Characterization_of_polycapillaries}).

For most polycapillaries $T_{max}$ is rather stable, and this is what should be expected for parallel structures.~\cite{Wolff2009}
However, for some polycapillaries the transmission first rises and then drops with energy. 
This effect can be attributed to 
a deflection from parallel structure geometry and is normally observed for curved polycapillaries where X-ray radiation is always transmitted in a series of bounces.~\cite{Wolff2011}
The transmission drop in higher energies is mostly caused by the decrease of the total reflection angle.
But, as was already presented in Figure~\ref{fig:reflectivity}, the increased number of bounces in the structure also decreases the transmission probability.
This effect is even more pronounced for lower energies. 
Accordingly, for a magnifying optic the parallel beam propagates without bounces only in a limited number of channels; the others are approached at a certain angle.  
This is also the case for imperfect parallel  optics for which the channels are slightly bended similarly as in a curved polycapillary.

%Such dependence was expected as a result of equation \eqref{eq:phi_c}.
%Note that, % $\alpha_{in}$ is always larger that the critical angle $\phi_c$ of total reflection for the glass material.
Even though all the polycapillaries are made of the same glass material, the values of angular acceptance $\alpha_{in}$ differ significantly from one objective to another.
Nevertheless, for all the optics $\alpha_{in}$ is inversely proportional to photon energy \eqref{eq:phi_c},
and follows the trend of $\varphi_c$ -- critical angle for total reflection.
Note also that, except for the 8:1/2 optic,  the angular acceptance is always below $\varphi_c$.  
Optic 8:1/2 exhibits the largest angular acceptance.
This is an effect of conical geometry, that, according to equation \eqref{eq:alpha_in}, systematically raises the acceptance by $\alpha/2$ -- the half of conical aperture of a polycapillary channel.
For  parallel optics a clear trend is that lower $d_{in}/l$ ratio
%narrower channel diameter 
results in a lower angular acceptance.   
For thin and long channels the number of bounces is much higher that for the wider and longer ones.
This decreases the probability of a photon transmission.
For example an \unit{8}{keV} X-ray beam entering the 1:1/7 optic  at $\varphi_c$ would be reflected on the inner wall of a channel over 200 times; for the case of  1:1/60 optic the beam would encounter less than 30 bounces.

The optical sensitivity $S$ does not follow the trend of the parallel beam transmission $T_{max}$.
$S$ is  proportional rather to a product of $T_{max}$ and $\alpha_{in}^2$. 
As a result, for the optics with a relatively low transmission but high enough acceptance angle, $S$ can be still significant.
For 1:1/60 optic an increase of $S$ can be observed for X-ray energies above \unit{20}{keV}.
In this regime the optic becomes transparent to X-rays and photons begin to penetrate trough the facets not being absorbed.
  
Note that $S$, similarly to $\alpha_{in}$, decreases with energy. 
Consequently, the best optic sensitivity is expected for soft X-rays.

 \subsection{Resolution}

 \begin{table*}[t]
 	{
 		\small
 		
 		\footnotesize
 		\begin{tabular}{@{\extracolsep{\fill}}l@{}cl  rrrr l rrrr c @{\extracolsep{-1.5em}}r@{\extracolsep{\fill}} }
 			&&& 	\multicolumn{9}{c}{ measured  }  & calc. \\
 			\cmidrule{4-12}\cmidrule(lr){13-13}
 			%	\noalign{\smallskip}
 			\multirow{2}{*}{Optic}                       &         \multirow{2}{*}{\parbox{.5cm}{\centering $f_1$ [mm]}    }      & \multirow{2}{*}{ROI\ }     &     \multicolumn{4}{c}{$R^{20\%}$ [$\mu$m]  }     &  &     \multicolumn{4}{c}{$R^{10\%}$ [$\mu$m]  }     &
 			\multirow{2}{*}{\parbox{.9cm}{\centering $R_{PSF}/M$  [$\mu$m] }  }  \\ \cmidrule{4-7}\cmidrule{9-12}
 			%\noalign{\smallskip}
 			&                     &   & 1$\times$1 & 2$\times$2 & 3$\times$3 & 4$\times$4 &  & 1$\times$1 & 2$\times$2 & 3$\times$3 & 4$\times$4 &  &  \\ % \hline\noalign{\smallskip}
 			
 			\rowcolor{Gray}	\multicolumn{3}{r}{$p$	: } & 48\phantom{.0} & 24\phantom{.0} & 16\phantom{.0} & 12\phantom{.0} & & 48\phantom{.0} & 24\phantom{.0} & 16\phantom{.0} & 12\phantom{.0} & \multicolumn{2}{r}{$r_{NS}$ :}\\ 
 			\multirow{2}{*}{1:1/20{\it (3)}} &      \multirow{2}{*}{10}      & Sn L & 96.7       &       94.1 &       93.4 &       93.4 &  & 76.2       &       61.0 &       60.0 &       60.0 & ---                                     &  \\
 			&                               & Cu K & 73.8       &       67.3 &       66.9 &       66.6 &  & 62.1       &       53.3 &       51.3 &       49.7 & 48.9                                    &  \\
 			\multirow{2}{*}{1:1/20{\it (7)$^*$}} & \multirow{2}{*}{\phantom{0}7} & Sn L & 70.3       &       64.4 &       62.3 &       62.3 &  & 59.3       &       50.3 &       46.2 &       45.8 & ---                                     &  \\
 			&                               & Cu K & 63.1       &       53.1 &       53.4 &       53.1 &  & 51.4       &       44.5 &       44.1 &       44.5 & 42.5                                    &  \\
 			\multirow{2}{*}{1:1/7}       &      \multirow{2}{*}{14}      & Sn L & 78.0       &       68.6 &       67.0 &       67.0 &  & 63.0       &       53.2 &       52.3 &       52.6 & ---                                     &  \\
 			&                               & Cu K & 62.4       &       53.1 &       52.1 &       52.1 &  & 53.0       &       45.9 &       43.9 &       43.6 & 41.4                                    &  \\
 			%\cline{2-12}	\noalign{\smallskip}
 			\rowcolor{Gray}	
 			\multicolumn{3}{r}{$p$	: } & 6\phantom{.0} & 3\phantom{.0} & 2\phantom{.0} & 1.5 & & 6\phantom{.0} & 3\phantom{.0} & 2\phantom{.0} & 1.5 & &\multicolumn{1}{r}{$r_{NS}$ :}\\ 
 			\multirow{2}{*}{8:1/2}       &     \multirow{2}{*}{$<$1}     & Sn L & 8.0        &        7.0 &        7.0 &        6.6 &  & 6.6        &        5.6 &        5.6 &        5.6 & ---                                     &  \\
 			&                               & Cu K & 6.5        &        5.0 &        4.8 &        4.6 &  & 5.2        &        4.0 &        4.0 &        3.9 & \phantom{0}3.7                          &
 		\end{tabular}% 
 		%\begin{tabular}{@{\extracolsep{\fill}}  rr l rr c @{\extracolsep{\fill}} }
 		%	\multicolumn{2}{c}{$r^{20\%}$ [LP/mm]  } &  & \multicolumn{2}{c}{$r^{20\%}$ [LP/mm]  } &  \multicolumn{1}{c}{ $r_{PSF}$}   \\\cline{1-2}\cline{4-5}
 		%		\noalign{\smallskip}
 		%	1$\times$1 &                  4$\times$4 &  & 1$\times$1 &                  4$\times$4 & \multicolumn{1}{c}{   [LP/mm]  }    \\ \hline\noalign{\smallskip}
 		%	5.2        &                         5.4 &  & 6.6        &                         8.3 & --- \\
 		%	6.8        &                         7.5 &  & 8.1        &                        10.1 & \phantom{0}10.5 \\
 		%	7.1        &                         8.0 &  & 8.4        &                        10.9 & --- \\
 		%	7.9        &                         9.4 &  & 9.7        &                        11.2 & \phantom{0}12.5 \\
 		%	6.4        &                         7.5 &  & 7.9        &                         9.5 & --- \\
 		%	8.0        &                         9.6 &  & 9.4        &                        11.5 & \phantom{0}14.2 \\
 		%	62.2       &                        76.1 &  & 75.6       &                        89.1 & --- \\
 		%	76.6       &                       109.6 &  & 95.6       &                       128.4 & 131.5
 		%\end{tabular} 
 		\begin{tabular}{@{\extracolsep{\fill}}  rrrr l rrrr c@{} }
 			\multicolumn{9}{c}{ measured  }  & calc. \\
 			\cmidrule{1-9}\cmidrule(lr){10-10}
 			\multicolumn{4}{c}{$r^{20\%}$ [LP/mm]  }      &  &     \multicolumn{4}{c}{$r^{10\%}$ [LP/mm]  }      & \multirow{2}{*}{\parbox{1cm}{ \centering $r_{PSF}$  [LP/mm]} } \\
 			\cmidrule{1-4}\cmidrule{6-9}
 			%	\noalign{\smallskip}
 			1$\times$1 & 2$\times$2 & 3$\times$3 & 4$\times$4 &  & 1$\times$1 & 2$\times$2 & 3$\times$3 & 4$\times$4 &      \\
 			\rowcolor{Gray}	 10.4 & 20.8 & 31.2 & 41.7 & & 10.4 & 20.8 & 31.2 & 41.7 & % \hline\noalign{\smallskip}
 			\\
 			5.2        &        5.3 &        5.4 &        5.4 &  & 6.6        &        8.2 &        8.3 &        8.3 &  ---\\
 			6.8        &        7.4 &        7.5 &        7.5 &  & 8.1        &        9.4 &        9.7 &       10.1 &  \phantom{0}10.2 \\
 			7.1        &        7.8 &        8.0 &        8.0 &  & 8.4        &        9.9 &       10.8 &       10.9 &  ---\\
 			7.9        &        9.4 &        9.4 &        9.4 &  & 9.7        &       11.2 &       11.3 &       11.2 &  \phantom{0}11.7 \\
 			6.4        &        7.3 &        7.5 &        7.5 &  & 7.9        &        9.4 &        9.6 &        9.5 &  ---\\
 			8.0        &        9.4 &        9.6 &        9.6 &  & 9.4        &       10.9 &       11.4 &       11.5 &  \phantom{0}12.1 \\
 			%\noalign{\smallskip}	
 			\rowcolor{Gray}83\phantom{.0} & 167\phantom{.0} & 250\phantom{.0} & 333\phantom{.0} & &83\phantom{.0} & 167\phantom{.0} & 250\phantom{.0} & 333\phantom{.0} &	\\
 			62.2       &       71.1 &       71.6 &       76.1 &  & 75.6       &       88.5 &       88.5 &       89.1 &  ---\\
 			76.6       &      100\phantom{.0} &      104\phantom{.0} &      110\phantom{.0} &  & 95.6       &      126\phantom{.0} &      124\phantom{.0} &      128.4 &  136\phantom{.0}
 		\end{tabular} 
 		
 		$^*$ Measurements with 1:1/20{\it (7)} optic were performed at the HS-PIXE beam line at HZDR.
 	}
 	\caption{
 		Measured and calculated resolution of several polycapillary optics (installed at various sample-optic distances $f_1$)  for two energy Regions Of Interest (ROI) corresponding to Sn L  and  Cu K lines series.
 		Resolution is given as resolution limits $R$ and as resolution frequencies $r$.   
 		Values obtained from the measured 
 		Siemens Star gray disc estimation
 		with two cut off contrast levels of 10\% and 20\% are listed for 1$\times$1,  2$\times$2, 3$\times$3, and 4$\times$4 subpixel divisions.	 		
 		The subpixel sizes $p$ and corresponding Nyquist-Shannon frequencies $r_{NS}$ are  shown for reference.   
 		For  Cu K lines series the measured values are also compared to theoretical values calculated from the size of Point Spread Function (PSF) with equation \eqref{eq:R_PSF}. 
 	}
 	\label{tab:resolution}
 \end{table*}

 \begin{figure*}[b]
 	\centering
 	\hfill\hfill
 	\includegraphics[width=\linewidth]{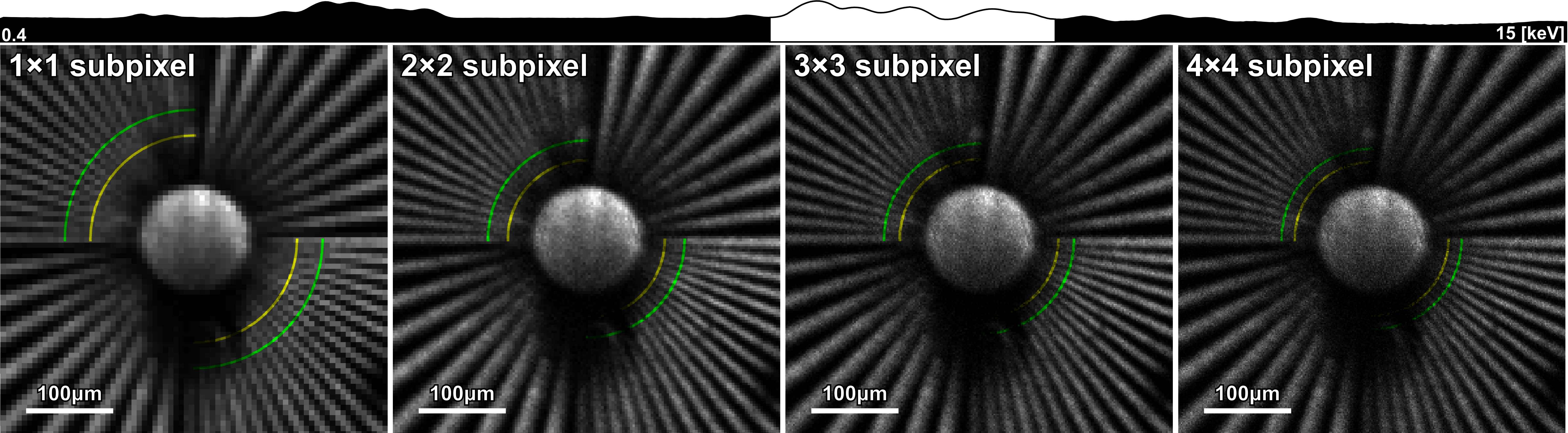}
 	% 	\includegraphics[width=.24\linewidth]{1x1_map__}\hfill
 	%	\includegraphics[width=.24\linewidth]{2x2_map__}\hfill
 	%	\includegraphics[width=.24\linewidth]{3x3_map__}\hfill
 	%	\includegraphics[width=.24\linewidth]{4x4_map__}
 	%\hfill 1$\times$1 subpixels \hfill\hfill 2$\times$2 subpixels \hfill\hfill 4$\times$4 subpixels \hfill~
 	\caption{
 		Cu K  intensity image of a Siemens star measured with  the 8:1/2 magnifying optic and rendered for $1\times1$, $2\times2$, $3\times3$, and $4\times4$ subpixel divisions.
 		Green and yellow rings indicate the gray disc borders for, respectively, 20\% and 10\% contrast levels.
 		On top a logarithmic representation of a sum spectrum of the full-image area is presented with the energy region corresponding to the Cu K lines series indicated in withe.
 	}
 	\label{fig:Siemens_CuK}
 \end{figure*}

The resolution of the \SLcam was evaluated for different types of optics and for two energy regions: the Sn L line series  from \unit{2.8\ to\ 4.6}{keV}, and the Cu K line series  from \unit{7.5\ to\ 10.3}{keV}. 
The use of these two energy ranges was possible due to the specific structure of the Siemens star that employs a thin Cu layer to improve the adhesion of the Sn elements.
The assessment of the resolution was performed for 1x1, 2x2, 3x3 and 4x4 subpixel divisions of acquired images. 
The cutoff contrast level was specified as 10\% and 20\%.

Table \ref{tab:resolution} lists the obtained measures of the resolution given as resolving power $R$ and as the line pairs frequency $r$. 
For  the Cu K line series these quantities are compared to  theoretical values: 
the size of a point spread function $R_{PSF}$, and 
the  corresponding resolution limit $r_{PSF} = 1/(2R_{PSF})$. 
Theoretical values were calculated  using equation \eqref{eq:R_PSF}   employing  parameters from Table \ref{tab:params} and Figure \ref{fig:characterisation_curves} (for focusing  8:1/2 optic $\alpha_{out}$ was measured to be \unit{\sim 0.05}{\degree}; for all the optics the distance to the detector is $f_2$\unit{ = 8}{mm}).
%; calculations for 1:1/20{\it(3)} an 1:1/20{\it(7)} optics were performed with the same parameters).
For comparison the table also gives the size of a single subpixel and the Nyquist-Shannon frequency $r_{NS}$ for corresponding subpixel divisions.

In most cases a $2\times2$ subpixel division gives already very good  results. 
Further pixel division only slightly improves the resolution.  
The choice of the contrast level significantly affects the obtainable resolution. 
In all the cases the resolution limit going from $R^{20\%}$ to $R^{10\%}$ is improved by $\sim20\%$
and closely approaches the theoretical limit  of $R_{PSF}/M$ when subpixel resolution is applied. 
These results are  illustrated in Figure~\ref{fig:Siemens_CuK}.

\begin{figure*}[t!]
	\centering
	\includegraphics[width=.75\linewidth]{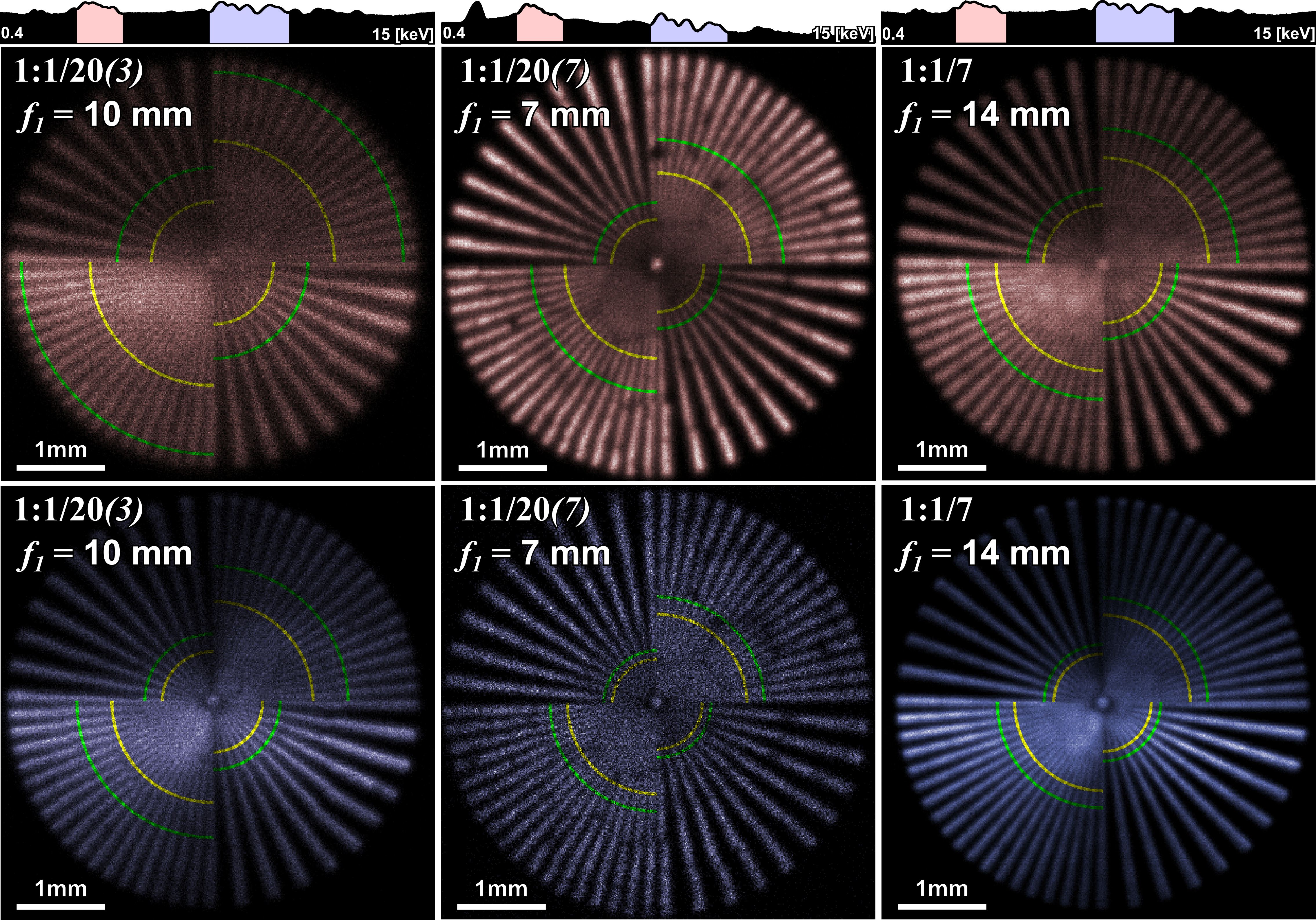}
	%\includegraphics[width=.25\linewidth]{Sn_12} 
	%\includegraphics[width=.25\linewidth]{Cu_12}\\
	%\includegraphics[width=.25\linewidth]{Sn_7}
	%\includegraphics[width=.25\linewidth]{Cu_7}
	%\includegraphics[width=.49\linewidth]{4x4_map_Sn_}\hfill
	%\includegraphics[width=.49\linewidth]{4x4_map__}
	%\hfill 1$\times$1 subpixels \hfill\hfill 2$\times$2 subpixels \hfill\hfill 4$\times$4 subpixels \hfill~
	\caption{
		Images of the Siemens star measured with 1:1 polycapillary optics.
		Images for  different sample-optic distances and for two energy regions, corresponding to   Sn L  and  Cu K lines series, are presented.
		On top logarithmic representations of the sum spectra of the full-image area are shown with indicated corresponding energy regions.
		Images are rendered with $4\times4$ subpixel division.
		Green and yellow rings indicate the gray disc borders for  20\% and 10\% contrast levels respectively.
	    The  images employing 1:1/20{\it (7)} optic were measured  at the HS-PIXE beam line at HZDR. 
	}
	\label{fig:Siemens_4x4}
\end{figure*}

A very important aspect when using polycapillary optics is the strong energy dependence of resolution.
According to equation \eqref{eq:R_PSF} the resolving power of a polycapillary should scale with the acceptance angle which is inversely proportional to energy. 
As a result the  images are  subject to chromatic aberration -- a better resolving power is expected for higher X-ray energies.
This trend was observed for all the polycapillaries --
the resolution limits achieved for Cu K line series are always better then the ones obtained for Sn L lines.  

The influence of the sample-optic distance $f_1$ on imaging is another effect that strongly influences the measurement. 
As stated in Section~\ref{sec:Resolving_power} the size of a point spread function is reduced when minimizing $f_1$.
As a result the resolving power is improved, but also the resolution variation with energy  is decreased.
This can be seen in Figure \ref{fig:Siemens_4x4} where  results obtained at various distances and for different energy regions are compared.

\subsection{Real sample example}
\begin{figure*}
	\centering
	\includegraphics[width=.7\linewidth]{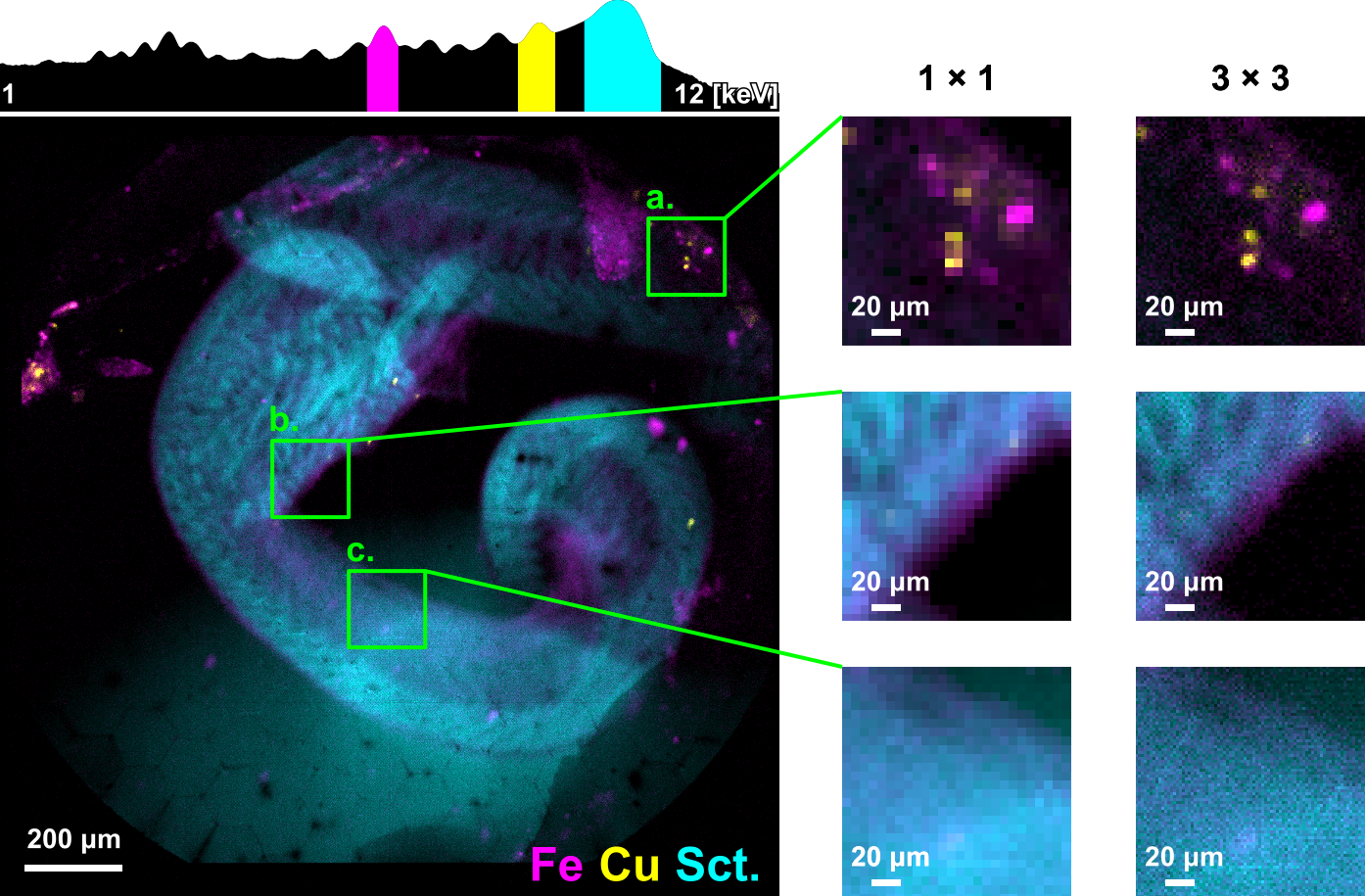}
	
	\caption{Image of a snail radula measured %at the BAM{\it line} at BESSY~II 
		 with the 8:1/2 magnifying lens.
		Magenta, yellow and cyan colors are assigned to the intensity of, respectively, Fe~K$\alpha$ and Cu~K$\alpha$ fluorescence
		lines, and scattered X-ray radiation.
		The main image is displayed with $3\times 3$ subpixel resolution.
		On top a logarithmic representation of a sum spectrum of the full-image area is presented.
		Panels on the right side of the image represent $\times 3$ magnified image details in standard (\unit{6}{\mu m } pixel size) and $3\times 3$ subpixel (\unit{2}{\mu m } pixel size) resolution.   }
	\label{fig:radula}
\end{figure*}

In order to present the color X-ray camera performance on a real sample a snail radula was measured with the 8:1/2 focusing optic using synchrotron light set to the energy of 10 keV.
The  false color image presented in Fig.~\ref{fig:radula}  was
obtained by superposing color maps corresponding to the intensity
of scattered X-ray radiation and the characteristic fluorescence lines of Fe and Cu.
The main image represents the whole available \SLcam chip  area of $264\times 264$ pixels and was rendered in $3\times 3$ subpixel resolution.
In the insets smaller details of the image are compared for standard and subpixel resolution.

As expected the  subpixel resolution makes it possible to resolve small overlapping elements.
With the optic used it was possible to resolve several micrometer spots at the energies of Fe K$\alpha$ and Cu K$\alpha$ lines (see inset a.). 

In the presented image the scattered radiation follows the pattern of the Fe K$\alpha$ image.
As a result the chromatic aberration can be clearly seen -- the lower energy image is more blurred and creates a border fringe around the higher energy image -- this is illustrated in the inset b.

Finally, the polycapillary optic defects (see inset c.) can serve as a measure of the lowest resolution limit that could be obtained with the optic. 
The image of the optic's defects does not result from the imaging object, instead it is created within a polycapillary.
For that reason $f_1$ should be set to $0$ in order to evaluate its resolution  with equation \eqref{eq:R_PSF}.
The resulting resolving power limit is  \unit{2.6}{\mu m}.
And indeed, hexagonal lines of width comparable with the subpixel size of \unit{2}{\mu m} can be resolved.

%\begin{figure}
%	\centering
%	\includegraphics[width=\linewidth]{one_droplet}
%	
%	\caption{   }
%	\label{fig:droplets}
%\end{figure}

%
%\begin{table}[hbtp]
%\begin{tabular}{ccccc}
%\toprule
%Optics & Spot & Subpixel & $R_{\upmu m}$ & $R_{LP/mm}$ \\
%$[\,name\,]$ & $[\,\upmu m\,]$ & $[\,\upmu m\,]$ & $[\,\upmu m\,]$ & $[\,$LP/$mm\,]$ \\
%\midrule
%20$\,\upmu m$ 1x1  & 37 & 48  &  64(6) & 8\\
%20$\,\upmu m$  3x3 & 37 & 16  &  36(3)  & 14\\
%20$\,\upmu m$  5x5 & 37 & 9.6 &  35(2)  & 14\\
%\midrule
%7$\,\upmu m$ 1x1    & 24  & 48  & 55(6) & 9\\
%7$\,\upmu m$  3x3   & 24  & 16  & 30(3)  & 17\\
%7$\,\upmu m$  5x5   & 24  & 9.6 & 27(2)  & 18\\
%\bottomrule 
%\end{tabular}
%\caption{Resolution obtained from the Siemens star analysis. The (sub)pixel size is calculated from the \unit{48}{\micro\meter} chip pixel size. The spot size is the capillary diameter + the divergence at \unit{6}{\milli\meter}, using \unit{0.2}{\degree} acceptance angle for the capillaries.}
%\end{table}

\section{Conclusions}

The resolution of a polycapillary optic is a function of many factors. 
Some of them, like sample to optics distance, or the energy region used for imaging can be adapted on a regular basis depending on a running experiment.
However, most of the resolution sensitive parameters are defined during the polycapillary fabrication process. 
This includes, but is not limited to, the channel diameter, the glass material used, and the  magnification factor.

The most relevant parameter for resolution is the channel entrance diameter $d_{in}$. 
The polycapillary resolving  power given by the Nyquis-Schannon theorem is about $2d_{in}$.  
Besides, $d_{in}$ is also the ultimate limit for resolving power resulting from the size of the point spread function.

The image resolution can be 
considerably  improved by reducing the sample-optic distance.
Accordingly, the best resolution would be achieved if the sample was positioned directly on the lens.
Such a setup could be realized, e.g., in reversed TXRF geometry where the polycapillary objective covered by a thin foil would serve as a reflector.  
Such a setup would prevent the detection of the primary beam radiation.

In practice such a radical shortening of the sample-optic distance would be effective only for the focusing optics.
In case of parallel optics the spacing effect is less dramatic, due to a more significant influence of the optic-detector gap, and a \unit{1}{mm} distance
is sufficient to obtain a maximum improvement of the resolution.
In  Table~\ref{tab:theoretical_res} the size of the PSF is compared for a sample-optic distance of \unit{1}{mm} and direct positioning on the objective. 
In case of a focusing optic the resolution limit can be decreased by a factor of 2; not much difference is encounter in case of parallel optics, however.
%It is worth noting that, for the given polycapillaries, such a close sample installation brings the size of PSF below the  Nyquist-Schannon theorem limitation.
Note that such a close sample installation brings the size of PSF below the resolution limit given by the Nyquist-Schannon theorem which is double the size of a capillary channel.
\begin{table}[h!]
	{\small
		\begin{tabular}{cccc}
			$f_1$\unit{}{[mm]} & \multicolumn{3}{c}{ $R_{PSF}/M$\unit{}{[ \mu m] } }  \\
			\rowcolor{Gray}      \multicolumn{1}{r}{Optic:}            & 1:1/20{\it(3)} & 1:1/7 &           8:1/2           \\
			1                           &  35.8  & 21.3  &            4.3            \\
			0                           &  35.7  & 21.1  &            2.7
		\end{tabular}\\ 
%		\begin{tabular}{cccc}
%			Optic          & $f_1$\unit{}{[mm]} & $R_{PSF}/M$\unit{}{[ \mu m] } &  \\ \hline
%			\multirow{2}{*}{1:1/20} &         1          &        35.8        &  \\
%			&         0          &        35.7        &  \\
%			\multirow{2}{*}{1:1/7}  &         1          &        21.3        &  \\
%			&         0          &        21.1        &  \\
%			\multirow{2}{*}{8:1/2}  &         1          &   \phantom{0}4.3   &  \\
%			&         0          &   \phantom{0}2.7   &
%		\end{tabular}
	}
	\caption{Theoretical limits to resolving power calculated with \eqref{eq:R_PSF} for Cu K line series. The results neglect the Nyquist-Schannon theorem limitation. }
		\label{tab:theoretical_res}
\end{table}

The spatial resolution varies with X-ray energy following the change in critical angle of total reflection $\varphi_c$.
As a consequence,  the resolution is improved for higher energies.
In addition to  the energy dependence, the value of $\varphi_c$ is strongly influenced  by the properties of the reflecting material. 
The higher the optical density of the material, the lower  $\varphi_c$ is. 
Accordingly, the size of a PSF, and consequently the resolution limit, can be seriously reduced if a glass with a high Pb concentration is used. 
However, a decrease of $\varphi_c$  entails a decrease of acceptance angle $\alpha_{in}$ and, as a result, the polycapillary sensitivity would be reduced.
Therefore the fabrication of polycapillary objectives needs to compromise the need for better resolution with good enough optics sensitivity.

In order to visualize the images with the resolution given by the polycapillary lens an appropriate pixel size is needed.
In this respect a subpixel resolution algorithm is a very productive tool.
However, the maximum number of subpixel divisions is restricted by the electronic properties of the detector (see Section~\ref{sec:CTF}). In addition the pixel division results in a lower number of counts in each virtual pixel; consequently, the available contrast is decreased.

Image resolution can be significantly improved when magnifying optics with a conical shape are employed.  
The conical optics, however, have high acceptance angles and can exhibit a halo.\cite{bjeoumikhov2005capillary}

With all that said, arriving at a resolving power of \unit{1}{\mu m} is challenging and would require at most a \unit{0.5}{\mu m} entrance channel. 
Research on a further reduction of the capillary channel diameter and halo minimization are pursued.
An objective with channel dimensions below \unit{1}{ \mu m} can be expected in the near feature.

%
%Polycapillary optics can be efficiently used for high spatial resolution of XRF and PIXE imaging. 
%
%

%
%Ix`
%
%
%
%The exit diameter of an imaging polycapillary optics used with \SLcam ranges between 7 and 24~$\mu$m.
%The pnCCD pixel dimension is \unit{48}{\mu m}. 
%Consequently, the resolution of a standard \SLcam image is limited by the pixel size.
%
%A straight polycapillary  optics used for 1:1 imaging  can provide a resolution of \unit{35}{\mu m}.
%With the use of conical shaped magnifying optics spatial resolution can be limited to several microns.
%Approaching \unit{1}{\mu m} is possible.
%In comparison to the scanning micro XRF method an overview qualitative elemental distribution is obtained within a few seconds and the spatial resolution can approach \unit{1}{ \mu m} using laboratory X-ray sources.
%

\section*{Authors Contribution}

SHN, MP, AB, ZB, JT, OS, and RW designed and planned the polycapillary characterization experiment.
AB and ZB provided the polycapillary optics.
JT created motorization and acquisition software.
SHN and MP performed the polycapillary characterization.
MR and UR realized imaging at synchrotron radiation facility.
SHN, JB, JvB, FM, ADR, and OS carried out PIXE measurements.
SHN and OS provided the image rendering and analyzing software. 
SHN, MP, JB, and OS handled the data.
SHN  analyzed the data, carried out the calculations and wrote the manuscript. 
All the authors reviewed the manuscript.

\section*{Acknowledgments}

We thank Björn Stelbrink and Andreas Wessel (Museum für Naturkunde Berlin) for providing the snail radula sample.
This work has been supported by Marie Curie Actions - Initial Training Networks (ITN) as an Integrating Activity Supporting Postgraduate Research with Internships in Industry and Training Excellence (SPRITE) under EC contract no.\ 317169.
SHN acknowledges financial support from  the Swiss National Science Foundation (SNSF), Project no.\ 148569.

%\begin{figure}
%\centering
%\includegraphics[width=1\linewidth]{./html/Images2_08}
%\caption{Intensity profile across Au lines structure  calculated for normal resolution (black line) and 3x3 subpixel division (red line) images. 
%Estimated positions of the Au stripe centers are indicated with black vertical lines. 
%The intensity profile was constructed from data sampled at regular intervals across the Au lines. 
%To increase the signal to noise ratio the intensity was integrated over the pixelated plane along the whole length of Au lines. 
%}
%\label{fig:Au_stripes_profile}
%\end{figure}

%\begin{figure}
%\centering
%\includegraphics[width=1\linewidth]{./html/Images2_09}
%\caption{Contrast Transfer Function (MTF) calculated for subpixel (red line) and normal (blue line) resolution images  of the Au stripes structure. }
%\label{fig:Au_stripes_profile}
%\end{figure}

%\begin{figure*}
%\centering
%\includegraphics[width=.5\linewidth]{./Cr_structure1}%
%\includegraphics[width=.5\linewidth]{./Cr_structure2}
%\caption{}
%\label{fig:Cr_structure}
%\end{figure*}
%
%\begin{figure*}
%\centering
%\includegraphics[width=.45\linewidth]{./snail_1x1}\hfill
%\includegraphics[width=.45\linewidth]{./snail_3x3}
%\caption{}
%\label{fig:snail_1x1}
%\end{figure*}

\footnotesize{
%\bibliography{D:/Dokumenty/PRACA/papiery/bibliografia}
\bibliography{bibliografia_pub}
\bibliographystyle{rsc} %the RSC's .bst file
\inputencoding{utf8}
}

\end{document}